# A Lossless Intra Reference Block Recompression Scheme for Bandwidth Reduction in HEVC-IBC

Jiyuan Hu[1], Jun Wang[2,3,4 ✉], Guangyu Zhong[1], Jian Cao[1], Ren Mao[5], Fan Liang[1,6]
[1]*School of Electronics and Information Technology, Sun Yat-sen University, Guangzhou, China*
[2]*School of Microelectronics Science and Technology, Sun Yat-Sen University, Zhuhai, China*
[3]*Southern Marine Science and Engineering Guangdong Laboratory (Zhuhai), Zhuhai, China*
[4]*ZhuHai JieLi Technology Co., Ltd., Zhuhai, China*
[5]*ZhuHai South IC Design Service Center, Zhuhai, China*
[6]*Peng Cheng Laboratory, Shenzhen, China*
{hujy23, zhonggy5, caoj33}@mail2.sysu.edu.cn, maor89@163.com, {wangj387, isslf}@mail.sysu.edu.cn

*Abstract*—The reference frame memory accesses in inter prediction result in high DRAM bandwidth requirement and power consumption. This problem is more intensive by the adoption of intra block copy (IBC), a new coding tool in the screen content coding (SCC) extension to High Efficiency Video Coding (HEVC). In this paper, we propose a lossless recompression scheme that compresses the reference blocks in intra prediction, i.e., intra block copy, before storing them into DRAM to alleviate this problem. The proposal performs pixel-wise texture analysis with an edge-based adaptive prediction method yet no signaling for direction in bitstreams, thus achieves a high gain for compression. Experimental results demonstrate that the proposed scheme shows a 72% data reduction rate on average, which solves the memory bandwidth problem.

## I. Introduction

During the inter prediction of multimedia applications such as HEVC, massive data traffic between the video codec and frame memory is inevitable because of the reading and writing of original and reconstructed frames. Off-chip memory DRAM is used to buffer those frames for its relatively lower price compared to on-chip memory. Memory bandwidth requirements ascend rapidly with the increasing demand for higher definition videos and the increasing complexity of video coding standards. As a result, DRAM consumes more than half of the entire system's power, the bandwidth of DRAM is a critical bottleneck in ultra high definition codec design [1]. Therefore, it is essential to minimize the DRAM bandwidth requirement in inter prediction.

On the other hand, intra prediction suffers the same bandwidth problem. With the emerging screen content applications such as online meetings, screen content coding (SCC) is extended to HEVC [2]. Intra block copy [3] is a new kind of intra prediction and performs efficiently in SCC. The previously reconstructed block with a similar pattern in the current frame is utilized as the predictor by IBC as shown in Fig. 1. Typically samples after reconstruction and filtering operations are stored into off-chip DRAM, and loaded into on-chip memory for motion estimation and compensation in inter prediction. When the IBC mode is involved, there is an additional memory bandwidth requirement for writing the unfiltered reconstructed samples of the current picture to DRAM for motion estimation and compensation in IBC. As a result, the memory bandwidth is more intensive by the adoption of IBC.

To the best of our knowledge, the recompression in intra prediction method IBC has not been discussed, however, the recompression algorithms in publications for inter prediction are still valuable for reference.

Inter reference frame recompression (Inter-RFRC) is a technique to reduce the memory bandwidth by compressing the reference frame data to be stored in the off-chip memory DRAM. The bandwidth requirement is reduced because the amount of fetched and stored data is reduced.

Inter-RFRC algorithms are mainly divided into two groups: lossy and lossless. Generally, compared to lossless Inter-RFRC, lossy Inter-RFRC [4-6] can attain a higher and fixed data reduction rate. As a result, it inherently supports random access for each coded block and saves memory size requirements. Lossy Inter-RFRC is carried out in the core video coding loop in both encoder and decoder to avoid the mismatch and the loss of the reference frame will be compensated by residuals. However, the degradation in coding efficiency limits its range of application.

Lossless Inter-RFRC algorithms usually contain three parts: prediction, entropy coding, and memory organization part.

For the prediction part, the algorithms fall into two categories: block-wise prediction and pixel-wise adaptive prediction. A simple block-wise prediction is horizontal direction differential pulse-code modulation (DPCM) scanning used in [7], where the current sample is predicted by the value of the left neighboring sample. [8] uses both horizontal and vertical direction DPCM with a signaling bit coded to indicate the direction. Intra Mode Referenced In-Block Prediction [4] utilizes eight modes to get better prediction accuracy and needs to code the signaling bits too. However, the texture information is not paid attention to in [7, 8] and intra prediction mode information is adopted to help the mode decision process in [4], which makes it can only be used on the encoder side unless removing its intra mode mapping. As for the pixel-wise adaptive prediction, it utilizes the texture information of each sample to make a mode decision, thus no signaling bit is required. Self-Contained Directional Intra Prediction is proposed in [9], which uses the above 2x2 samples as well as the left 2x2 samples to calculate the edge direction of the current sample, and one of seven candidate samples is selected by the edge direction. Gradient-adjusted prediction (GAP) in CALIC [10] and median edge detector (MED) in JPEG-LS [11] are two adaptive predictors. GAP uses 7 samples to model the texture information and MED uses 3 samples. [12] compares the performance of GAP and MED in embedded compression.

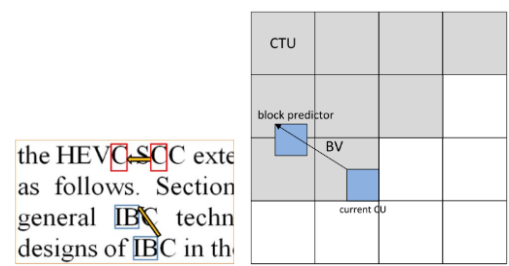

Fig. 1. IBC in HEVC-SCC [3]

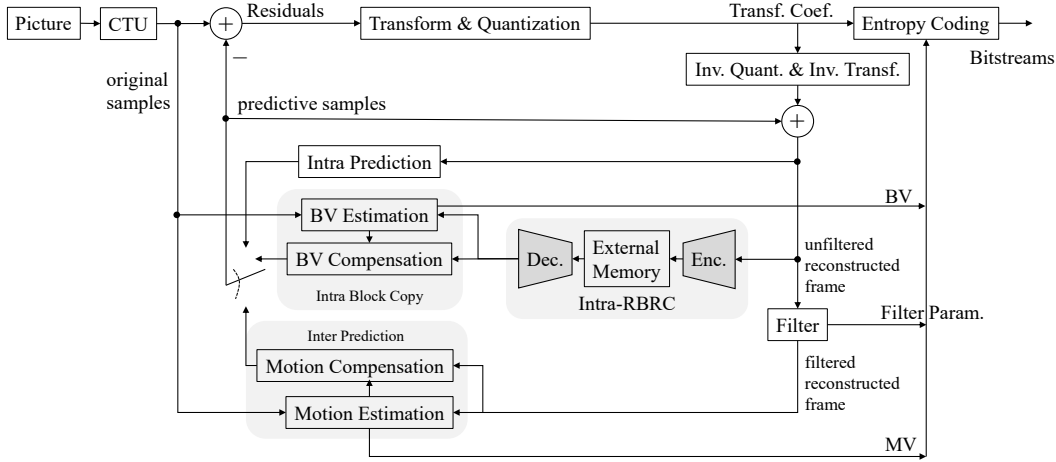

Fig. 2. Video coding architecture with Intra Reference Block Recompression.

For the entropy coding part, Adaptive Golomb-Rice coding [13, 14], Dynamic Exp-Golomb coding [9], static Huffman coding [15], Significant Bit Truncation (SBT) Coding [16] and Semi-fixed-length Code Table [7], Small-value Optimized Variable Length Coding (SVO-VLC) Table [4], etc., are different attempts to eliminate the statistics redundancy.

And for the memory organization part, fixed addressing [7, 16] is of low complexity, while variable addressing [4, 9] further improves the DRAM efficiency by aligning the storage of coded blocks. However, address table storage overheads, as well as a complex memory controller, is required in variable addressing.

To reduce the memory bandwidth, we proposed an intra reference block recompression (Intra-RBRC) scheme on top of Inter-RFRC. Moreover, a lossless edge-based recompression scheme is proposed and outperforms previous ones in terms of data reduction rate.

The remainder of this paper is organized as follows. The proposed reference block recompression scheme, especially the edge-based adaptive prediction, is explained in Section II. Section III shows the experimental results of the proposed scheme. The conclusion is given in Section IV.

## II. PROPOSED SCHEME

Inspired by previous Inter-RFRC algorithms used in inter prediction, we extend the Inter-RFRC to IBC in intra prediction, named Intra Reference Block Recompression (Intra-RBRC). Fig. 2 shows a conceptual block diagram describing the video coding architecture with Intra-RBRC. Intra-RBRC compresses the unfiltered reconstructed blocks used in IBC before storing them into external memory DRAM and decompresses the compressed data during BV estimation and compensation, thus the data traffic, i.e., memory bandwidth in IBC is reduced.

### A. Edge-based Adaptive Prediction

The frame is partitioned into 8x8 blocks, whose size is discussed in Section II. C. The first sample of the block is directly stored without compression. The first row and column of the block are coded by horizontal and vertical DPCM, which use the left and above sample respectively as the predictor and calculate the residual by subtracting the original sample with the predictor. The rest samples of the block are predicted by our proposed edge-based prediction.

As shown in Fig. 3. (a), to reduce the computation complexity, the neighboring 4 samples named $r_1$, $r_2$, $r_3$, and $r_4$ is used as the reference samples to predict the current sample x. In some special cases, x is in the right column of a block, $r_4$ are unavailable and thus set equal to $r_3$. Our goal is to choose the best one reference sample $r_n$ as the predictor, which is closest to x in value.

We proposed an effective method called edge-based adaptive prediction which takes the texture information into account. The edge direction is obtained by calculating the gradient near the current sample x, then the reference samples choosing problem turns into edge direction problem as shown in Fig. 3. (b). We define the edge direction's horizontal component as $Dx$ and vertical component as $Dy$ as shown in Fig. 3. (c). As a result, from the figure, we can easily discover that $r_4$ is the best predictor when $Dy$ is approximately equal to $Dx$, $r_3$ is the best predictor when $Dy$ is far more than $Dx$, and so on. The full representation shows in (1). According to (1), both encoder and decoder can attain their prediction without extra signaling bits. To be more hardware-friendly, we use shift operation (<< and >>) rather than multiplication and division. $s$ is the sign of the direction. $\oplus$ means "exclusive or", i.e., when $Dx$ and $Dy$ have the same sign, $s$ equals 0.

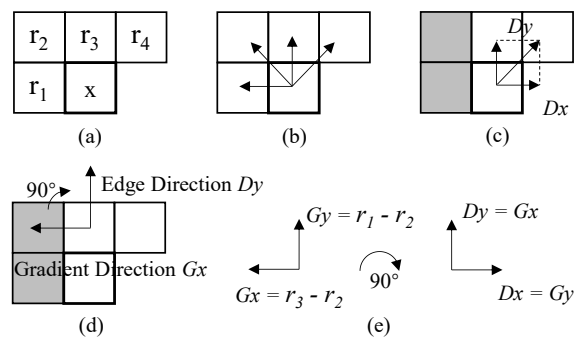

Fig. 3. Illustration of Edge-based Adaptive Prediction

$$x = \begin{cases} r_4, & (|Dx| >> 1) < |Dy| \leq (|Dx| << 1) \ \& \ s = 0 \\ r_3, & |Dy| > (|Dx| << 1) \\ r_2, & (|Dx| >> 1) < |Dy| \leq (|Dx| << 1) \ \& \ s = 1 \\ r_1, & |Dy| \leq (|Dx| >> 1) \end{cases} \quad (1)$$

where $s = dy \oplus dx$

Edges are significant changes of intensity in an image. Gradient indicates the intensity change. Edge direction is perpendicular to the direction of maximum intensity change (gradient direction) as shown in Fig. 3. (d). Therefore, we know the edge's vertical component $Dy$ equals horizontal gradient $Gx$ and horizontal component $Dx$ equals vertical gradient $Gy$ as shown in Fig. 3. (e).

To reduce the computational complexity when deriving the gradient, the Roberts Operator (2) [17] is concerned and we modify it into (3) since the right bottom sample is the unavailable current sample x to be predicted.

$$Mx = \begin{bmatrix} -1 & 0 \\ 0 & 1 \end{bmatrix}, My = \begin{bmatrix} 0 & -1 \\ 1 & 0 \end{bmatrix} \quad (2)$$

$$Mx = \begin{bmatrix} -1 & 1 \\ 0 & 0 \end{bmatrix}, My = \begin{bmatrix} -1 & 0 \\ 1 & 0 \end{bmatrix} \quad (3)$$

$$Img = \begin{bmatrix} r_2 & r_3 \\ r_1 & x \end{bmatrix} \quad (4)$$

Therefore, the edge direction's horizontal component $Dx$ and vertical component $Dy$ is calculated by the convolution of the image (4) and the operator (3) as shown in (5).

$$\begin{cases} Dy = Gx = Img \otimes Mx = r_3 - r_2 \\ Dx = Gy = Img \otimes My = r_1 - r_2 \end{cases} \quad (5)$$

Fig. 4 shows an example of a 4x4 luma block in the screen content test sequence "FlyingGraphics" [18] using our proposed edge-based adaptive prediction. The number shows the value of each sample. The "$r_n$" under each number is the predictor derived by our proposal. The arrows represent the direction of the predictor. This example shows the proposed prediction method can model the texture information well and attain an accurate prediction direction.

There are two merits in our proposal. Firstly, compared to block-wise prediction, our proposal takes the texture information into account to make an accurate pixel-wise

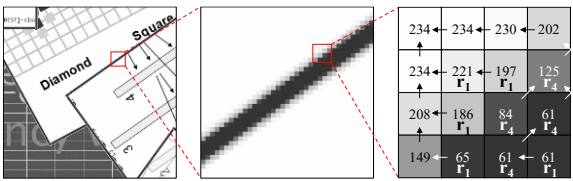

Fig. 4. An example of Edge-based Adaptive Prediction

TABLE I. SMALL-VALUE OPTIMIZED VLC TABLE

| Tab | 00 | 01 | 10 | 110 | 1110 | 11110 | 111110 | 111111 |
|---|---|---|---|---|---|---|---|---|
| Max | 0 | 1 | 2 | 3~4 | 5~8 | 9~16 | 17~32 | >32 |
| 0 | - | 1 | 01 | 001 | 0001 | 00001 | 00001 | |
| ±1 | | 0S | 1S | 01S | 001S | 0001S | 0001S | |
| ±2 | | | 00S | 10S | 010S | 0010S | 0010S | |
| ±3 | | | | 11S | 011S | 0011S | 0011S | |
| ±4 | | | | 000S | 100S | 0100S | 0100S | |
| … | | | | | … | … | … | |
| ±7 | | | | | 111S | 0111S | 0111S | |
| ±8 | | | | | 0000S | 1000S | 1000S | |
| … | | | | | | … | … | |
| ±11 | | | | | | 1011S | 1011S | xx…xS |
| ±12 | | | | | | 1100S | 11000S | |
| … | | | | | | … | … | |
| ±15 | | | | | | 1111S | 11011S | |
| ±16 | | | | | | 00000S | 1110000S | |
| … | | | | | | | … | |
| ±31 | | | | | | | 1111111S | |
| ±32 | | | | | | | 0000000S | |

S indicates the sign of the residual.

adaptive prediction. Secondly, the prediction direction is derived by the calculation (1) on both encoder and decoder, hence it saves the extra signaling bits indicating the prediction direction. A high gain is achieved by these two merits.

### B. Variable Length Coding

The residuals after the Edge-based Adaptive Prediction are small values compared to original samples. During the entropy coding process, the small-value optimized variable length coding (SVO-VLC) table proposed in [4] and shown in Table I is used since it exploits the coding potential of small value residual blocks.

The first sample in each residual block is excluded since it is uncompressed. Each block is divided and coded in the unit of $4 \times 4$. The max absolute value of each unit is used to choose a suitable table. Every residual in the same unit shares the same VLC table which is chosen by the max absolute value of each unit. When the max value is greater than 32, the residuals are coded by the 0-th order Exp-Golomb Coding. The table code and residual code are encoded into the bitstream.

Since without quantization process, the proposed Intra-RBRC scheme is lossless, as a result, there is no degradation in video quality.

### C. Block Size Selection

There is a trade-off between the data reduction rate (DRR) and the hardware complexity. The experimental results in Table II reveal that generally a bigger block size results in a better data reduction rate. However, a bigger block size is with higher probability to read unnecessary data and the hardware cost also increases as the block size increases. Although DRR of the $16 \times 16$ block shows the best DRR value among variable block sizes, the DRR increase from $8 \times 8$ to $16 \times 16$ is approximately 2% which is much smaller than the DRR increase from $4 \times 4$ to $8 \times 8$ that is 10%. Therefore, we choose $8 \times 8$ as our Intra-RBRC block size.

### D. Memory Organization

Because of the variable length of the compressed block data, random accessibility can be a challenge since a block's start address is hard to attain during hardware implementation. To solve this problem as well as simplifying the complexity of the memory controller, we use a low complexity memory organization method that stores the compressed data in fixed addresses which is the same as original data as shown in Fig. 5.

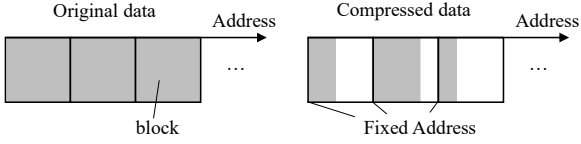

Fig. 5. The memory organization.

## III. EXPERIMENTAL RESULTS

The data reduction rate (DRR) shows how much data can be saved after compression. The definition of DRR is:

$$DRR = 1 - \frac{compressed\ data\ size}{original\ data\ size} \quad (6)$$

Like many other designs [5, 16, 19], the bandwidth reduction varies depending on the specific codec system design such as data alignment and reuse, thus it is essentially estimated as data reduction.

The experiment is carried on IBC scenario, all the 13 YUV420 test sequences, including text and graphics with motion (TGM), mixed content (MC), animation (ANI) category, specified in HEVC common test conditions for SCC [18]. Since the lossless recompression does not bring any quality loss, the proposed scheme is transparent to the codec, thus we extract the unfiltered and filtered reconstruction frame from the HM reference software (HM-16.21+SCM-8.8 [20]) under the encoder_intra_main_scc configuration (where QP=32) to conduct our test.

To evaluate the efficiency of the proposed Intra-RBRC scheme, especially the prediction part, six previous prediction works are also simulated on SVO-VLC, including horizontal DPCM prediction (HD) in [7], both horizontal and vertical DPCM prediction (HVD) with one extra signaling bit encoded in [8], a modified version of In-Block Prediction (4IBP) [4] that use the frequently used mode 0, 4, 5, and 6 instead of the intra mode mapping, Self-Contained Directional Intra Prediction (DIP) in [9], gradient-adjusted prediction (GAP) in [10, 12], and median edge detector (MED) in [11, 12]. The simulation results on the unfiltered reconstruction frame used by IBC are shown in Table III.

Table III reveals that both the proposed scheme and MED prediction [11, 12] can reach the highest DRR of 72%. Compared with the DIP [9] and GAP [10, 12], the proposed scheme achieves a better DRR with lower complexity since seven neighboring samples are utilized to model the content in DIP and GAP, while our proposal takes only three samples.

In addition, the proposed Intra-RBRC scheme can also be used in inter prediction. The DRR performance has been tested as shown in Table IV, where the intra scene is carried on unfiltered reconstruction sequences used by IBC while the inter scene is carried on the filtered reconstruction sequences used by inter prediction. The results show that the DRR degradation is about 2% when performing Intra-RBRC on inter prediction, which is acceptable.

## IV. CONCLUSION

In this paper, an efficient intra reference block recompression (Intra-RBRC) scheme for IBC in HEVC is proposed to reduce the memory bandwidth. With the merits of an accurate pixel-wise prediction and no signaling for direction, the proposed edge-based adaptive prediction in our Intra-RBRC scheme achieves a considerable data reduction rate of 72% on the intra IBC scene. Therefore, the proposed scheme can be efficient in solving the DRAM bandwidth and power problem. Future work will be high throughput hardware implementation to meet the real-time constraint in case of a video codec for high definition video sequences.

TABLE II. THE RELATION OF BLOCK SIZE AND DRR

| Block Size | | HD[7] | HVD[8] | 4IBP[4] | DIP[9] | GAP[10,12] | MED[11,12] | Proposed |
|---|---|---|---|---|---|---|---|---|
| 4x4 | Average | 60.0% | 58.5% | 56.8% | 61.1% | 60.1% | 61.7% | 61.6% |
| 8x8 | Average | 69.1% | 68.8% | 68.2% | 71.6% | 68.8% | 72.4% | 72.1% |
| 16x16 | Average | 71.4% | 71.4% | 70.9% | 74.6% | 70.7% | 75.6% | 75.1% |

TABLE III. DRR ON SCC SEQUENCES (UNFILTERED RECONSTRUCTION FRAME, QP=32)

| Class | Sequence | HD[7] | HVD[8] | 4IBP[4] | DIP[9] | GAP[10,12] | MED[11,12] | Proposed |
|---|---|---|---|---|---|---|---|---|
| TGM | FlyingGraphics | 65.2% | 65.0% | 64.4% | 68.2% | 65.2% | 68.2% | 68.1% |
| | Desktop | 61.4% | 61.2% | 60.4% | 67.6% | 63.2% | 69.0% | 69.0% |
| | Console | 69.5% | 69.2% | 68.5% | 74.9% | 69.8% | 76.1% | 76.3% |
| | ChineseEditing | 55.6% | 55.4% | 54.6% | 59.5% | 55.4% | 61.1% | 60.8% |
| | Web_browsing | 79.3% | 78.9% | 78.5% | 81.1% | 79.7% | 81.6% | 81.4% |
| | Map | 62.5% | 62.2% | 61.7% | 64.8% | 62.7% | 65.1% | 64.7% |
| | Programming | 69.2% | 68.8% | 68.2% | 72.4% | 68.5% | 73.3% | 73.1% |
| | SlideShow | 76.9% | 76.6% | 76.1% | 79.2% | 76.0% | 79.8% | 79.6% |
| MC | Basketball_Screen | 72.7% | 72.4% | 71.8% | 75.2% | 72.0% | 75.6% | 75.3% |
| | MissionControlClip2 | 72.9% | 72.6% | 72.0% | 75.3% | 73.2% | 75.9% | 75.6% |
| | MissionControlClip3 | 66.9% | 66.8% | 65.9% | 70.0% | 67.5% | 70.9% | 70.5% |
| ANI | Robot | 66.9% | 66.6% | 66.2% | 68.6% | 66.2% | 69.7% | 69.2% |
| | ChinaSpeed | 71.1% | 70.8% | 70.1% | 72.1% | 69.5% | 72.6% | 72.3% |
| **TGM** | **Average** | **67.5%** | **67.2%** | **66.5%** | **71.0%** | **67.6%** | **71.8%** | **71.6%** |
| **MC** | **Average** | **70.8%** | **70.6%** | **69.9%** | **73.5%** | **70.9%** | **74.1%** | **73.8%** |
| **ANI** | **Average** | **69.0%** | **68.7%** | **68.1%** | **70.3%** | **67.8%** | **71.2%** | **70.8%** |
| **Overall** | **Average** | **69.1%** | **68.8%** | **68.2%** | **71.6%** | **68.8%** | **72.4%** | **72.1%** |

TABLE IV. THE DRR ON DIFFERENT SCENE

| Scene | | HD[7] | HVD[8] | 4IBP[4] | DIP[9] | GAP[10,12] | MED[11,12] | Proposed |
|---|---|---|---|---|---|---|---|---|
| Intra | Average | 69.1% | 68.8% | 68.2% | 71.6% | 68.8% | 72.4% | 72.1% |
| Inter | Average | 67.7% | 67.3% | 66.7% | 69.9% | 67.4% | 70.9% | 70.5% |

ACKNOWLEDGMENT

This work is supported by Key R&D Program of Guangdong Province (No. 2019B010135002).